\pdfoutput=1
\documentclass[prd,preprint,eqsecnum,nofootinbib,amsmath,amssymb,preprintnumbers,tightenlines,longbibliography]{revtex4-1}

\usepackage{afterpage}
\usepackage{mathrsfs}
\usepackage{graphicx}
\usepackage{bm}
\usepackage{paralist}
\usepackage[colorlinks,linkcolor=blue,citecolor=blue]{hyperref}

\def\e{{\varepsilon} }
\def\E{{\mathcal E} }

\def\Re{{\rm Re}}

\def\ls2{{\ell_s^2}}

\newcommand\bga{\begin{align}}
\newcommand\nda{\end{align}}

\def\p{{\bm p}}

\def\Re{{\rm Re}}

\def\st{\begin{equation}}
\def\stp{\end{equation}}
\def\bg{\begin{eqnarray}}
\def\nd{\end{eqnarray}}
\def\Eq#1{eq.~(\ref{#1})}

\def\Fig#1{Fig.~\ref{#1}}
\def\Sect#1{Section~\ref{#1}}
\def\Ref#1{Ref.~\cite{#1}}

\def\llangle{\left\langle}
\def\rrangle{\right\rangle}

\def\drangle{\rrangle\!\rrangle}
\def\dlangle{\llangle\!\llangle}
\def\Bigdlangle{\Big\langle\!\!\Big\langle}
\def\Bigdrangle{\Big\rangle\!\!\Big\rangle}

\def\nott#1{\setbox0=\hbox{$#1$}                
   \dimen0=\wd0                                 
   \setbox1=\hbox{/} \dimen1=\wd1               
   \ifdim\dimen0>\dimen1                        
      \rlap{\hbox to \dimen0{\hfil/\hfil}}      
      #1                                        
   \else                                        
      \rlap{\hbox to \dimen1{\hfil$#1$\hfil}}   
      /                                         
   \fi}                                         %

\advance\parskip 1.9pt
\advance\voffset -0.2in

\def\st{\begin{equation}}
\def\stp{\end{equation}}
\def\bg{\begin{eqnarray}}
\def\nd{\end{eqnarray}}

\def\np{{\, .}}

\begin{document}

\title{Plane correlations and hydrodynamic simulations of heavy ion collisions}

\author{D.~Teaney}
\email{derek.teaney@stonybrook.edu}
\affiliation
    {%
    Department of Physics \& Astronomy,
    Stony Brook University,
    Stony Brook, NY 11794, USA
    }%
\author{L.~Yan}
   \affiliation
{
   CNRS, 
   Institut de Physique Th\'eorique de Saclay, F-91191
   Gif-sur-Yvette, France
}
 
\email{li.yan@cea.fr}

\begin{abstract}

   We use a nonlinear response formalism  to describe the event plane correlations measured by the ATLAS collaboration.  With one exception ($\left\langle \cos(2\Psi_2 - 6\Psi_3 + 4 \Psi_4) \right\rangle$), the event plane correlations are qualitatively reproduced by considering the linear and quadratic response to the lowest cumulants.  For the lowest harmonics such as $\left\langle \cos(2\Psi_2+3\Psi_3 - 5\Psi_5) \right\rangle$, the correlations are quantitatively reproduced, even when the naive Glauber model prediction has the wrong sign relative to experiment.  The quantitative agreement for the higher plane correlations (especially those involving $\Psi_6$) is not as good.  The centrality dependence of the correlations is naturally explained as an average of the linear and quadratic response.
\end{abstract}

\date{\today}

\maketitle

\section{Introduction}

The collective expansion of the deconfined fireball created in high energy
heavy-ion collisions maps the initial state of the Quark-Gluon Plasma (QGP)
to the final state particle spectrum.
The measured correlations in this spectrum
can  clarify the initial conditions and subsequent expansion dynamics of the  QGP \cite{Heinz:2013th,Hippolyte:2012yu,Teaney:2009qa}.  

On an event-event basis the azimuthal distribution
of produced particles can be decomposed into
a Fourier series
\st
\label{spectra_intro}
 \frac{dN}{ d\phi_\p  } 
= \frac{N}{2\pi}\,
\left(1 + 2\sum_{n=1}^{\infty} 
  v_n \cos(n \phi_\p - n\Psi_n)  
 \right) \, , 
\stp
and  the measured two particle correlation function determines the root mean square
of the these harmonics,  $\sqrt{\llangle v_n^2 \rrangle}$. 
The magnitude
of these harmonics is reasonably reproduced by event-by-event viscous
hydrodynamics provided the shear viscosity is not too large \cite{Heinz:2013th}. 
The correlations between the harmonics can provide new tests of the hydrodynamic
description, constrain the simulation parameters, and 
provide an estimate of the uncertainties in the computation. 
In this work we will describe the correlations between the observed event plane
angles $\Psi_n$ in order to clarify the 
expansion dynamics, and ultimately
to determine the shear viscosity of the QGP with credible systematic error bars.  

Clearly, an important input  to the hydrodynamic simulations is the distribution
of energy density in the transverse plane, which is usually estimated
from the known probability distribution of nucleons in the incoming nuclei.
There is reasonable evidence, both experimental \cite{Alver:2010gr} and theoretical \cite{Qiu:2011iv},  that $v_2$
and $v_3$ are to a good approximation linearly proportional to the
corresponding angular fluctuations in the  transverse energy density. 
However, event-by-event
hydrodynamic simulations have shown that the higher harmonics, $v_4$ and $v_5$, reflect
 both the response to 
corresponding angular harmonics in the initial state, and 
the non-linear hydrodynamic response which mixes lower order harmonics \cite{Qiu:2011iv,Gardim:2011xv}.
For example, the $5$-th flow harmonic, $v_5$, is determined in part by
the medium response to the $5$-th harmonic of the initial energy 
density distribution,  and in part by the non-linear mixing between $v_2$ and $v_3$.  
Such mode-mixing is especially important at high $p_T$ where
the non-linearities of the phase-space distribution play an important
role~\cite{Borghini:2005kd}. Indeed, there are indications that the dominant
source of mode mixing comes from freezeout as opposed to the hydrodynamic evolution~\cite{Floerchinger:2013tya}.
Motivated by these simulation results, and especially the simulation analysis of 
\Ref{Gardim:2011xv}, we  developed
a non-linear response formalism to describe the mixing between modes of
different order, and we investigated how the response coefficients depend
on centrality, shear viscosity, and transverse momentum \cite{Teaney:2012ke}.

These theoretical calculations preceded the corresponding 
experimental studies by the ATLAS \cite{ATLASCorrelations}  and ALICE collaborations \cite{Bilandzic:2012an}, which
qualitatively confirmed the mode mixing picture
by measuring significant
correlations between the event-plane angles of 
different orders\footnote{The ATLAS measurement did not precisely measure $\Psi_n$ \cite{Luzum:2012da}. Ultimately, this important first measurement will need to be redone, weighting the event averages with the $Q$ vector 
to provide an unambiguous quantity which can be compared to 
fairly  compared to simulations. See below for further discussion.}, {\it e.g.} between $\Psi_2$,$\Psi_3$, and $\Psi_5$. 
   Event-by-event hydrodynamics \cite{Qiu:2012uy} and AMPT calculations \cite{Bhalerao:2013ina} largely reproduce the structure of 
these correlations. 
The goal of this paper is to compare the response
formalism outlined in our previous work to the event-plane
correlations measured by the ATLAS collaboration \cite{Teaney:2012ke}.

As discussed more technically in \Sect{nlin_flow} we will use a non-linear response 
formalism 
to describe  the observed event plane correlations, 
rather than event-by-event hydrodynamics. 
In practice,
this means that we decompose the initial state into an average event
plus small fluctuations, which are systematically analyzed with cumulants.
The linear and quadratic response to each cumulant is found
by perturbing the average background, and finally
 the observed plane correlations are found by weighting the response functions
with the spectrum  of fluctuations.
Thus, the  response formulation provide a 
transparent link between the initial state and the final state, 
which contains {\it only} the linear and quadratic response  through a 
specified order in the cumulant expansion.
As we will see, this approach 
reproduces a lot of the observed event plane correlations, suggesting 
that most of the microscopic details of the initial state (beyond
the lowest cumulants) are irrelevant. Ideally, a limited number of 
initial state parameters can be extracted from experiment, and compared to
available theoretical frameworks such as the Color Glass Condensate
 to demonstrate the consistency and uniqueness of the approach.  There
 are indications that the spectrum of fluctuations from the Color Glass 
 Condensate is consistent with the observed harmonics \cite{Gale:2012rq}, but the uniqueness of this approach is not obvious.

A review of the non-linear flow response formalism will be given in
\Sect{nonlinear}.  This has several ingredients. First, the 
spectrum of initial fluctuations in various Glauber type models
is described in \Sect{cumulant},  and this spectrum is analyzed with the cumulant and  moment expansions. 
Then, we  describe how the response coefficients are calculated, and how these coefficients determine the plane correlations
in \Sect{nlin_flow} and \Sect{plane_formulation}.
Finally, we compare the response formalism to 
the ATLAS data in \Sect{diss} and discuss the results.

Throughout the paper $\Phi_n$  will denote
participant plane angle based  on the {\it cumulants} rather than moments.
(The correlations in the Glauber model between the cumulant angles $\Phi_n$ 
are markedly different from the correlations found using the analogous moment based
angles -- see \Sect{cumulant}.)
$\Psi_n$ denotes the event plane angle extracted from the final 
state momentum spectra.


\section{Review of nonlinear flow response formalism}
\label{nonlinear}


\subsection{Characterizing the initial state with cumulants}
\label{cumulant}

As discussed in the introduction, an important input to the hydrodynamic
calculations is the spectrum of initial fluctuations.  This spectrum is traditionally \cite{Alver:2010gr}
quantified with the  participant plane anisotropy based on moments\footnote{In this
formula we are using a moment based definition of $\e_n$ and $\Phi_n$. For most
of the text we will use a cumulant based definition.}
\st
\label{mon_def}
\varepsilon_n e^{in\Phi_n}\equiv- 
\frac{\llangle r^ne^{in\phi_r}\rrangle  }{\llangle r^n \rrangle}   \qquad \qquad \qquad {\mbox{(Not used).} }
\stp 
Here the brackets $\llangle\ldots \rrangle$  denote an average
over the participating nucleons of a single event,  while
$r e^{i\phi_r}=x+i\,y$ notates the transverse coordinates of the participants.  
It is convenient to use a 
complex notation $z\equiv x + iy$ so that $\e_n e^{in\Phi_n} = -\llangle z^n \rrangle /\llangle r^n \rrangle $.
As emphasized in our previous work,  it is often useful to 
characterize the fluctuations with cumulants rather than moments. The cumulants
subtract off the lower order correlation functions of $z$  to describe the
irreducible correlations
\st
\e_n e^{in\Phi_n}\equiv- \frac{1}{r^n} \left[ \llangle z^n \rrangle - {\rm subtractions  } \right] \, .
\stp
For example, the fourth order cumulant is
\st
\e_4 e^{i4\Phi_4}\equiv- \frac{1}{\llangle r^4 \rrangle} \left[ \llangle z^4 \rrangle - 
3 \llangle z^2 \rrangle^2 \right] \, , 
\stp
where the factor of three arises because there are three ways to pair four objects. 
Here and below we have  assumed that we are working in the center of mass 
coordinate system where $\llangle z \rrangle = 0$. 
The usefulness of cumulants can be understood by considering 
a Gaussian distribution, 
\st
\rho(x,y)  \propto e^{ - \frac{x^2}{2\llangle x^2 \rrangle }  - \frac{y^2}{2\llangle y^2 \rrangle } }  \,  ,
\stp
whose fourth order moment anisotropy $\llangle z^4 \rrangle$
is non-zero, and is trivially
correlated with the eccentricity, $\llangle z^2 \rrangle$.  
The fourth order cumulant takes out these
trivial correlations, and for a Gaussian distribution we have
$\e_4 \propto  \llangle z^4 \rrangle  -  3\llangle z^2 \rrangle = 0$.

\begin{figure}
\includegraphics[width=0.42\textwidth,height=0.9\textheight]{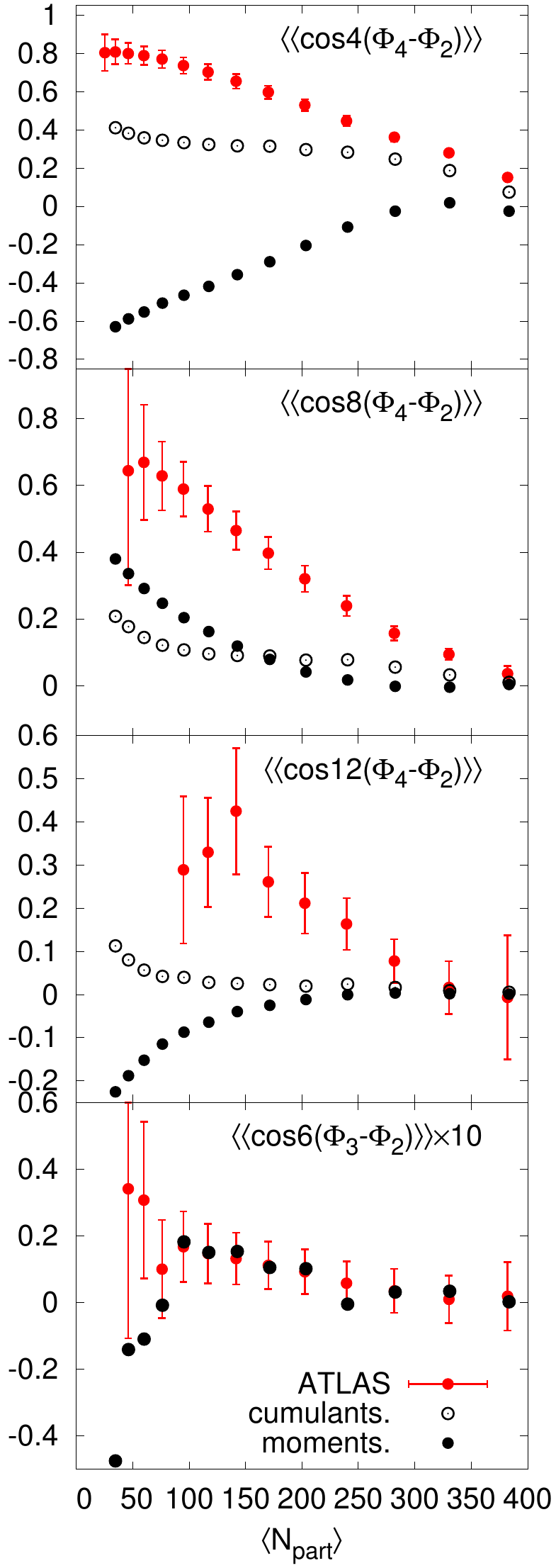}
\includegraphics[width=0.42\textwidth,height=0.9\textheight]{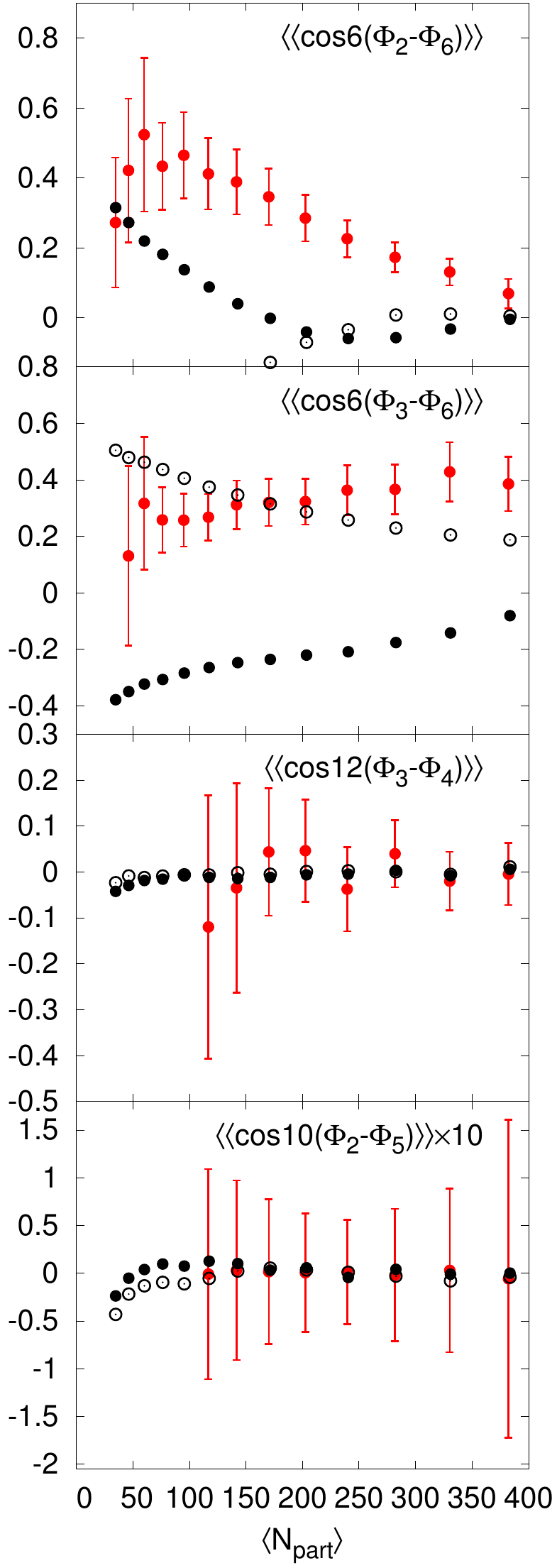}
\caption{ Participant 2-plane correlations from Phobos Monte Carlo Glauber model \cite{Alver:2008aq}
as measured  by  the cumulant and  moment expansions.  The measured 
event plane correlations \cite{ATLASCorrelations} are presented for reference and as a 
point of contact, and are not supposed to be directly compared to the 
Glauber model results. \label{ini_cor1}
}
\end{figure}

\begin{figure}
\includegraphics[width=0.49\textwidth]{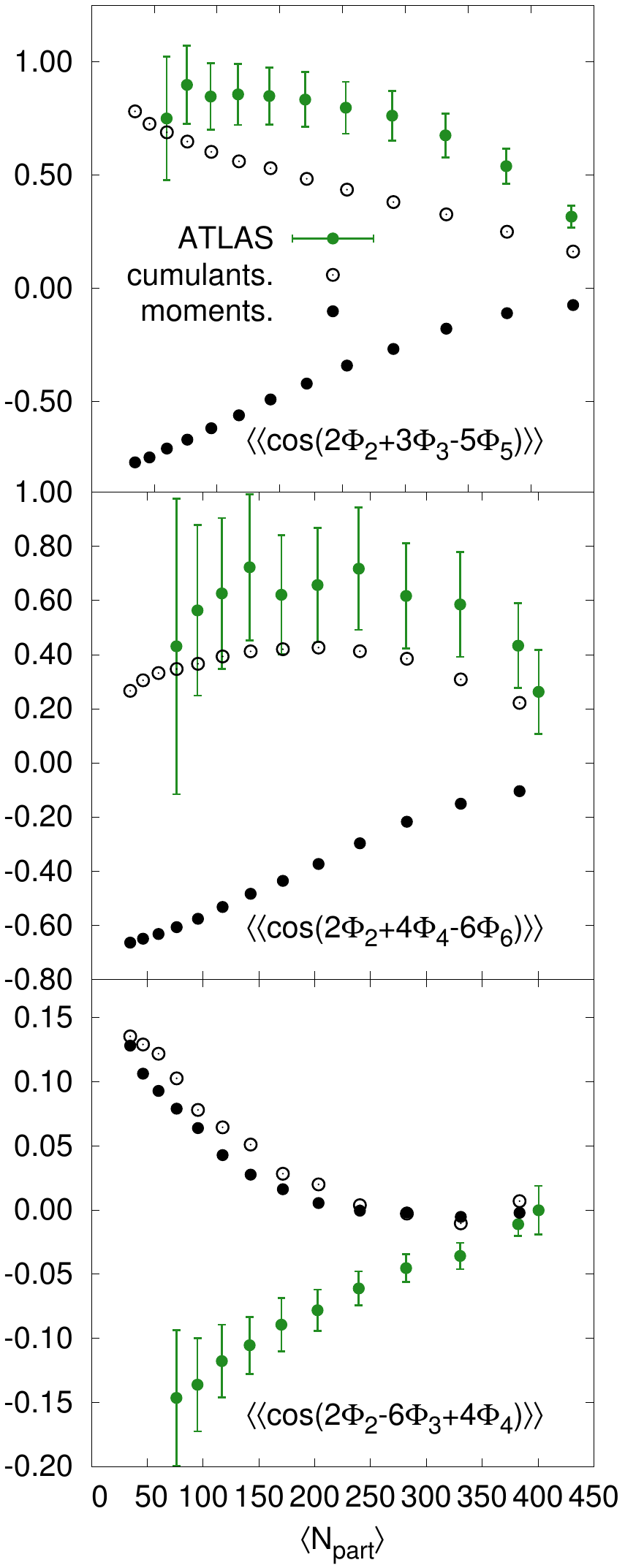}
\includegraphics[width=0.49\textwidth]{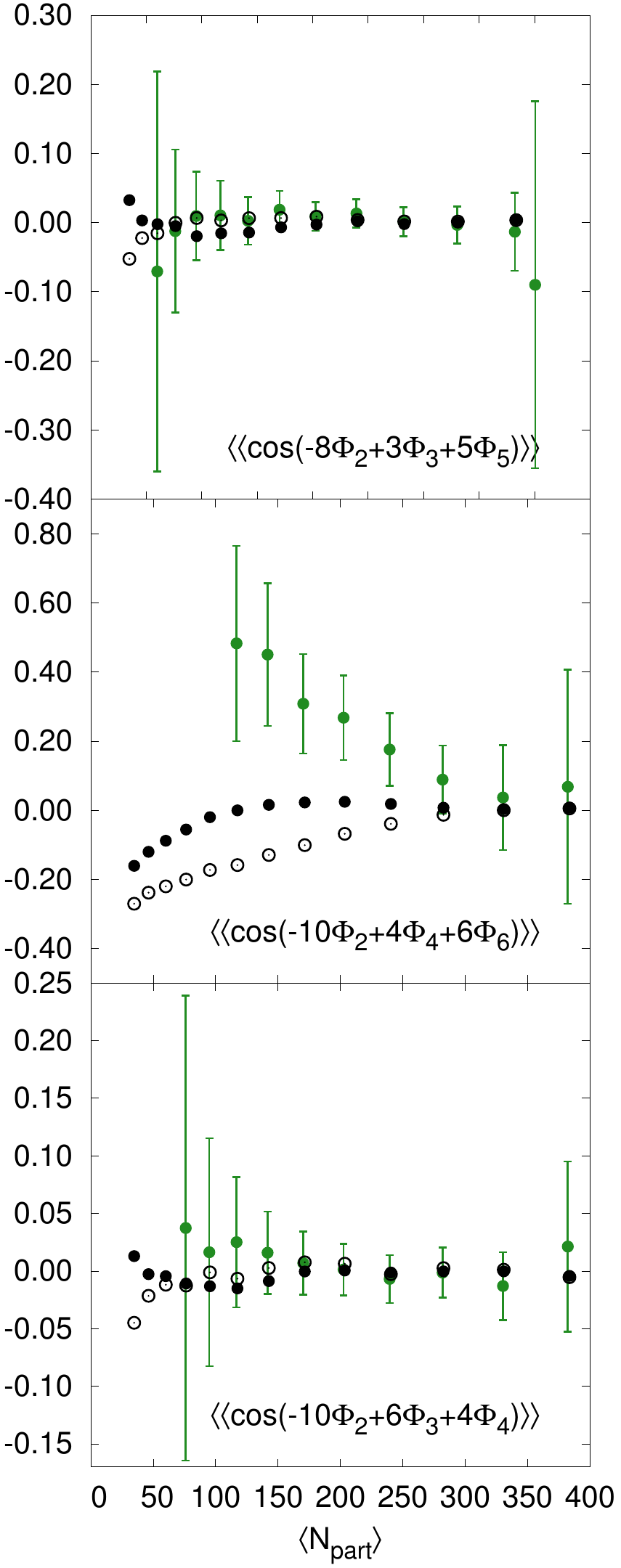}
\caption{ Participant 3-plane correlations from Phobos Monte Carlo Glauber model \cite{Alver:2008aq}
as measured  by  the cumulant and  moment expansions.  The measured 
event plane correlations \cite{ATLASCorrelations} are presented for reference and as a 
point of contact, and are not supposed to be directly compared to the 
Glauber model results.
\label{ini_cor2} 
}
\end{figure}

The azimuthal anisotropies through $\e_6$ are
\begin{align}
   \label{Endef}
   \E_2 \equiv \e_2 e^{i2\Phi_2} \equiv & -\frac{\llangle z^2 \rrangle}{\llangle r^2 \rrangle } \, ,   \\
   \E_3 \equiv \e_3 e^{i3\Phi_3} \equiv & -\frac{\llangle z^3 \rrangle}{\llangle r^3 \rrangle } \, ,   \\
   \E_4 \equiv \e_4 e^{i4\Phi_4} \equiv &- \frac{1}{\llangle r^4 \rrangle} \left[ \llangle z^4 \rrangle - 3 \llangle z^2 \rrangle^2 \right] \, ,  \\
   \E_5 \equiv \e_5 e^{i5\Phi_5} \equiv &- \frac{1}{\llangle r^5 \rrangle} \left[ \llangle z^5 \rrangle - 10 \llangle z^2 \rrangle \llangle z^3 \rrangle \right] \, ,  \\
   \E_6  \equiv \e_6e^{i6\Phi_6} =&  -\frac{1}{\llangle r^6 \rrangle }  \left[ \llangle z^6 \rrangle - 15\llangle z^4 \rrangle \llangle z^2 \rrangle  - 10 \llangle z^3 \rrangle^2 + 30 \llangle z^2 \rrangle^3  \right] \, ,
\end{align}
where $\E_n = \e_n e^{in\Phi_n}$ denotes the eccentricity and its phase.
The $\e_1$ which drives $v_1$ is a special case,  and is 
given by
\st
\E_1 = \e_1 e^{i\Phi_1 } \equiv  - \frac{1}{\llangle r^3 \rrangle } \llangle z^2 z^* \rrangle \, .
\stp
Given an initial state Glauber model for the distribution of nucleons 
such as Glissando \cite{Broniowski:2007nz} or the Phobos Monte Carlo Glauber model \cite{Alver:2008aq} 
one can calculate the correlations between the angles $\Phi_n$. 
\Fig{ini_cor1} and \Fig{ini_cor2} show such a calculation  from the Phobos 
Monte-Carlo model. Here and below the double brackets $\dlangle \ldots \drangle$ indicate an average over events, while the single brackets $\langle \ldots \rangle$ denote an average over one event.
In the Phobos Glauber  the participant centers are
used to define the averages in \Eq{Endef}, while in the Glissando model a 
slightly different prescription is used, which is based on the wounding profile of
the nucleon \cite{Broniowski:2007nz}. 

It is interesting to compare the correlations between the cumulant and moment
based angles.  For example, the  $\dlangle \cos4(\Phi_4 - \Phi_2) \drangle$ 
correlation is strongly negative with the moment based definitions, while the 
corresponding correlations with cumulant angles are positive.  
In the moment definition the $\Phi_4$,$\Phi_2$ correlation arises because the fourth order 
eccentricity $\E_4$ is trivially correlated with the second 
order eccentricity $\E_2$ through the average geometry. These
trivial geometric correlations are removed with the cumulant definition,
and the residual correlation is positive.
The correlation between $\Psi_4$ and $\Psi_2$ seen in the data is positive, but does not seem
to be directly related to the participant plane correlation between $\Phi_4$ and $\Phi_2$. The interpretation of the
data is described in \Sect{diss}. 

\subsection{Formulation of flow response}
\label{nlin_flow}
The harmonic flow $v_n$ and the corresponding flow angle $\Psi_n$ are defined by the Fourier decomposition of the final state  particle spectrum,
\st
\label{vndef}
\frac{dN}{d\phi_p}=\frac{N}{2\pi}\left[1+\sum_n\left(v_ne^{-in(\phi_p-\Psi_n)}+ c.c.\right)\right]\np
\stp
Here we use a complex expression, with {\it c.c.} standing for complex conjugate. For simplicity, we also define
a complex flow coefficient which takes into account the flow and its angle simultaneously, 
\st
V_n\equiv v_n e^{in\Psi_n}\np 
\stp

Following the same strategy and notation as in our previous work \cite{Teaney:2012ke}, 
the magnitude of the flow and its corresponding angle  is 
given by the  response  formula
\st
\label{Vn}
V_n=\left(\frac{w_n}{\varepsilon_n} \right) \E_n
+\sum_{\mbox{\tiny quadratic}} \left(\frac{w_{n(pq)}}{\varepsilon_p\varepsilon_q} \right)  \E_p \E_q + \ldots \, .
\stp
Here $w_n$ is the $n-$th linear response coefficient to 
a given $\E_n$, and $w_{n(pq)}$ are the $n-$th quadratic response coefficients.
The ellipses in \Eq{Vn} stand for higher order nonlinear contributions which 
are generally neglected in this work.  
The only exception  to this 
rule is for $V_6$ where we included the contribution from $\E_2^3$.
Even in this case, the $\E_2^3$ contribution 
was found to be numerically small compared to the
quadratic $\E_2\E_4 $ and the  $\E_3^2$ results. 
The current calculation uses the following minimal set of response coefficients
\st
\label{included}
w_1\ldots w_6 \qquad w_{1(32) }, \; w_{3(21)},  \;  w_{4(22) }, \; w_{5(23)}, \;  w_{6(24)}, w_{6(33)}, w_{6(222)} \, .
\stp
We found that additional non-linear terms such as $w_{2(31)}$, $w_{4(13)}$, and
$w_{5(14)}$ were not numerically important for the current set of correlations.
Thus, we reverted the code to the minimal set of response
coefficients listed in \Eq{included}. The effects of including
additional (radial) modes in the linear response was studied  
in \cite{Floerchinger:2013rya,Floerchinger:2013hza}. While a complete analysis will 
be presented in future work, a preliminary investigation shows 
that these (radial) contributions are small for the inclusive correlations
studied here.

The form of \Eq{Vn} indicates the dependence of the $n$-th order harmonic 
flow  and its angle  on the linear response coefficient $w_n$ and the quadratic response coefficients $w_{n(pq)}$. These response coefficients are
calculated by perturbing the (smooth) background geometry and 
determining the resulting flow. The details of this procedure 
have been given in our previous work \cite{Teaney:2012ke},
and here we will simply review the most important features.

Linear and nonlinear flow response coefficients are obtained from ``single-shot"
2+1D hydrodynamic simulations. In this approach the average geometry
for a given centrality class
is modeled with a cylindrically symmetric Gaussian, $i.e.$ the initial
entropy density in the event at Bjorken time $\tau_o$ is
\st
\label{Gauss}
s(x,y,\tau_o)  = \frac{C_s}{\tau_o \pi R^2}  e^{-r^2/R^2}   \, .
\stp
The rms radius of the Gaussian is adjusted to match the rms radius 
of a smooth (or averaged) Glauber model  for a given centrality. 
The overall constant of the Gaussian
is adjusted  as a function of centrality 
to reproduce the measured $dN_{\rm ch}/dy$ at the LHC~\cite{LiThesis}.
The response 
coefficients are calculated by perturbing this radially symmetric Gaussian by small deformations;
running the perturbed Gaussian through the  hydro tool chain; and finally calculating $w_n$ or $w_{n(pq)}$.  For example, for   we calculate
$w_5/\epsilon_5$ by deforming the Gaussian by a tiny  $\e_5$ and calculating $v_5$. Similarly we calculate $w_{5(23)}/(\e_2 \e_3)$ by deforming 
the Gaussian by $\e_2$ and by $\e_3$ and calculating  $v_5$, which is proportional  to $\e_2\e_3$.
To summarize, all of the response coefficients and their dependence on centrality  are obtained by simulating slightly deformed cylindrically symmetric Gaussian initial conditions.

We have implemented 2nd order BRSSS hydrodynamics, taking the 
necessary second order transport
coefficients from the AdS/CFT results. The numerical
scheme (but not the code) is similar to the scheme developed in \Ref{Dusling:2009df}.
The shear viscosity to entropy ratio $\eta/s$  
is constant throughout the whole evolution, 
and is set to the canonical  value of  $1/4\pi$.
We use
an equation of state that
parametrizes the lattice results \cite{Laine:2006cp}, which was used previously by Romatschke and Luzum \cite{Luzum:2008cw}. Finally, we use a constant freeze-out temperature $T_{\rm fo}=150$ MeV, and 
adopt
the widely used quadratic ansatz for the first viscous correction to the freeze-out distribution function \cite{Teaney:2013gca}.

\subsection{Formulation of plane correlations}
\label{plane_formulation}

The plane correlations are measured by event-plane method~\cite{ATLASCorrelations},
and a multi-particle correlation method~\cite{Bhalerao:2011bp,Bilandzic:2012an}. We will focus on the event 
plane method which was used by the ATLAS collaboration. The details
of this method were clarified by Luzum and Ollitrault who showed that if the 
event plane method is used, the  quantity that is measured depends
on the reaction plane resolution of the detector~\cite{Luzum:2012da}.     

We are interested in  
describing the correlations involving two and three event plane 
angles. For definiteness we will present formulas
for a specific correlation, 
$\dlangle \cos ( 4\Psi_{4} -  2(2\Psi_2) )  \drangle $,
which can be easily generalized to other harmonics.
(To aid the reader we have written $4\Psi_2=2(2\Psi_2)$ to expose the general pattern.)
The 4-2 plane correlation  is related to $V_4$ and $V_2$ through
\st
\label{24cor}
\dlangle\cos(4\Psi_4-2(2\Psi_2))\drangle=
\Bigdlangle \frac{ \Re\, (V_4 {V_2^*}^2)}{\sqrt{(V_4 V_4^*) (V_2 V_2^*)^2  } }  \Bigdrangle 
=\Bigdlangle\frac{w_4\cos4(\Phi_4-\Phi_2)+w_{4(22)}}{|w_4 e^{-i4\Phi_4}+w_{4(22)}e^{-i4\Phi_2}|}\Bigdrangle\, .
\stp
Thus, both the linear and nonlinear response coefficients enter this formula for the event plane correlation.

The ATLAS collaboration quantified the event plane correlations
by measuring related correlations 
between the experimental planes, $\hat \Psi_n$, as
determined by the $Q_n$-vectors, $\vec Q_n = |Q_n| e^{-in\hat \Psi_n}$ \cite{ATLASCorrelations}.
Further investigation showed that the measured quantity can not be directly
interpreted as an event plane correlation in the form of \Eq{24cor}. 
The measured correlation equals \Eq{24cor}
when the experimental event plane resolution approaches unity, 
\st
   \label{highres}
   \langle\cos(4\hat \Psi_2 - 2(2\hat \Psi_2))\rangle\{EP\} \simeq
   \Bigdlangle \frac{ \Re\, (V_4 {V_2^*}^2)}{\sqrt{(V_4 V_4^*) (V_2 V_2^*)^2  } }  \Bigdrangle \,  \qquad \mbox{(high resolution limit)}. 
   \stp
Here  we have notated the experimental quantity 
with $\{{\rm EP}\}$ \cite{ATLASCorrelations}, and refer to \Ref{Luzum:2012da} where the precise  definition is 
carefully examined.  
The notation for the experimental quantity is somewhat misleading since the experimental definition does not actually
correspond to the average of a cosine, and can be greater than 
one.
In the limit of low event plane resolution, the measured 
quantity equals
\st
\label{lowres}
\langle\cos(4\hat \Psi_2 - 2(2\hat \Psi_2))\rangle\{{\rm EP}\}
\simeq
\frac{ \dlangle \Re\, (V_4 {V_2^*}^2) \drangle } {
 \sqrt{\dlangle V_4 V_4^* \drangle
 \dlangle (V_2 V_2^*)^2 \drangle }  }\qquad {\mbox{(low resolution limit) } }.
\stp  
Clearly \Eq{lowres} differs from \Eq{highres} by how the events are 
weighted. 
The event plane measurements by the ATLAS collaboration (such as
$\langle\cos(4\hat \Psi_2 - 2(2\hat \Psi_2))\rangle\{{\rm EP}\}$)
interpolate
between the high and low resolution limits depending on the reaction plane resolution.

As the experimental resolution depends on the harmonic number, the 
detector acceptance, and centrality, we will compute  both the high
and low resolution limits and compare both curves
to the experimental data.  In the future,
such ambiguities in the measurement definition can be avoided
by measuring 
\st
\label{doneright}
\frac{\llangle v_4 v_2^2 \cos(4\hat \Psi_2 - 2(2\hat \Psi_2))\rrangle }
{ \sqrt{ \llangle v_2^2 \rrangle^2 \llangle v_4^2 \rrangle } }
= 
\frac{ \dlangle  \Re\, (V_4 {V_2^*}^2) \drangle } 
{ \sqrt{\dlangle V_4 V_4^* \drangle
\dlangle (V_2 V_2^*) \drangle^2 }   } \, , 
\stp
as originally suggested in~\cite{Bhalerao:2011yg},  and more recently in~\cite{Luzum:2012da}.
Such angular of correlations have already been measured by the ALICE
collaboration \cite{Bilandzic:2012an}, but we will not address this preliminary data here. 
Certainly \Eq{doneright} is the most natural from the perspective 
of the response formalism developed in this work.

Finally, we give
one additional example, $-8\Psi_2+3\Psi_3 +5\Psi_5$  of how a three plane correlation function 
is calculated  in the high and low resolution limits:
\st
   \langle\cos(-4(2\hat \Psi_2) + 3\hat \Psi_3 + 5\hat \Psi_5)\rangle\{{\rm EP}\}
   \simeq
\Bigdlangle 
\frac{\Re\,({V_2^*}^4 V_3 V_5)}{ 
\sqrt{ (V_2V_2^*)^4 (V_5 V_5^*) (V_3 V_3^*) }  }
 \Bigdrangle   \qquad \mbox{(high resolution)}, 
\stp
\st
\langle\cos(-4(2\hat \Psi_2) + 3\hat \Psi_3 + 5\hat \Psi_5)\rangle\{{\rm EP}\} 
\simeq
\frac{ \dlangle  \Re\,({V_2^*}^4 V_3 V_5 )\drangle } 
{ \sqrt{
\dlangle (V_2 V_2^*)^4 \drangle 
      \dlangle V_3 V_3^* \drangle
      \dlangle V_5 V_5^* \drangle
}   }  \qquad  \mbox{(low resolution)}.
      \stp
In the future the quantity which is most easily compared to 
theoretical calculations is 
\begin{align}
   \frac{\langle v_2^4v_3 v_5\cos(-4(2\hat \Psi_2) + 3\hat \Psi_3 - 5\hat \Psi_5)\rangle }{\sqrt{\llangle v_2^2 \rrangle^4 \llangle v_3^2 \rrangle \llangle v_5^2 \rrangle } }
 = 
\frac{ \dlangle  \Re\,({V_2^*}^4 V_3 V_5 )\drangle } 
{ \sqrt{
\dlangle V_2 V_2^*\drangle^4
      \dlangle V_3 V_3^* \drangle
      \dlangle V_5 V_5^* \drangle
}   }  \, .
\end{align}

\section{Discussions and conclusions}
\label{diss}

\begin{figure}
\begin{center}
\includegraphics[width=0.45\textwidth,height=0.9\textheight]{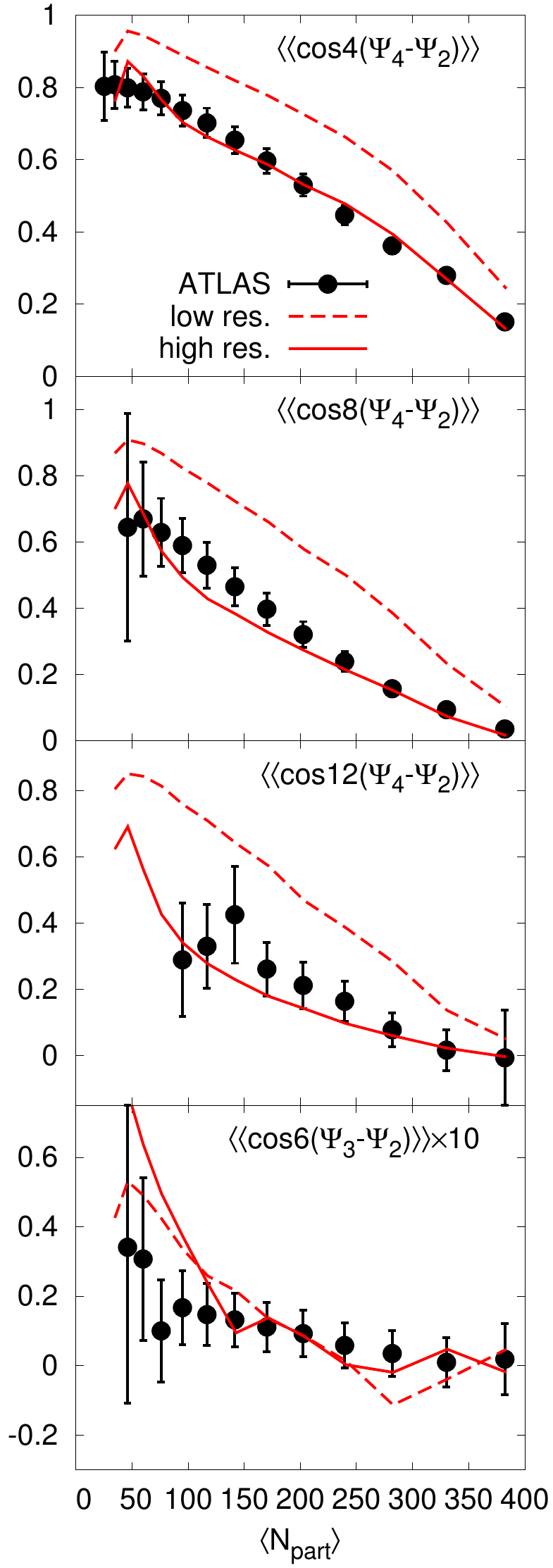} 
\includegraphics[width=0.45\textwidth,height=0.9\textheight]{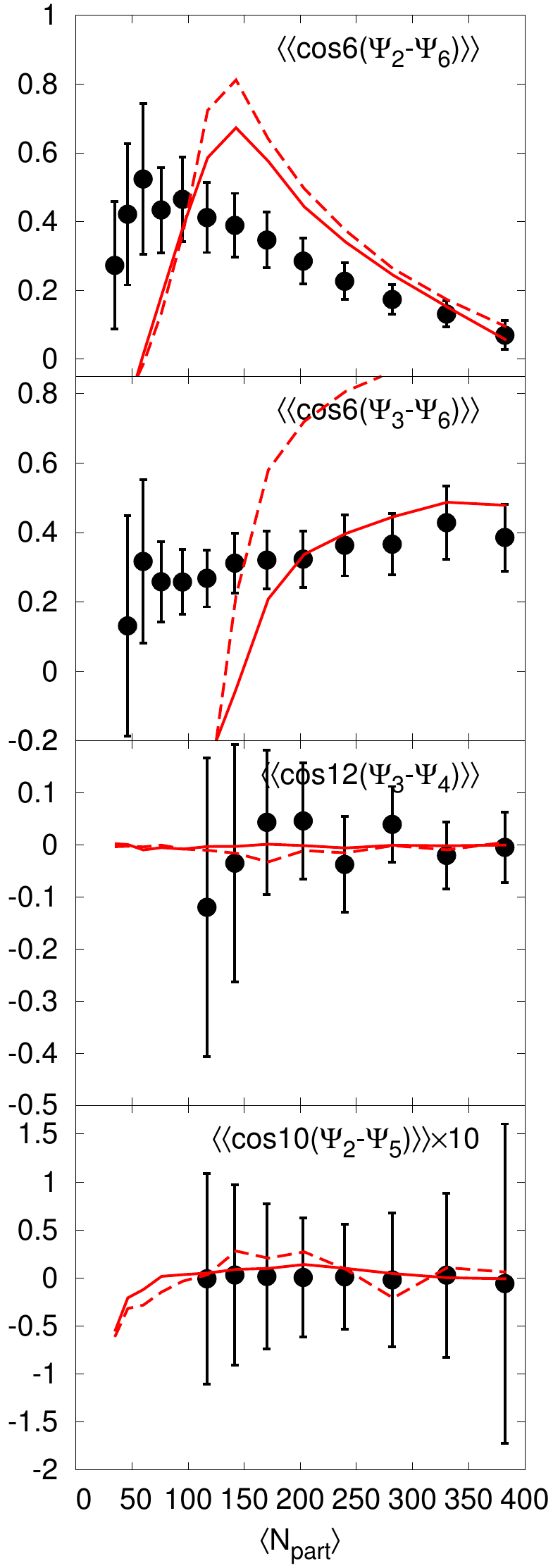} 
\caption{ 
Two plane correlations  using the non-linear response formalism. 
Here $\eta/s=1/4\pi$ for PHOBOS Monte-Carlo Glauber initial conditions.
The data are from the ATLAS collaboration \cite{ATLASCorrelations}.
The solid lines indicate the high resolution limit, \Eq{highres}, while the dashed
lines indicate the low resolution limit, \Eq{lowres}. 
\label{twoplanes}
}
\end{center}
\end{figure}

\begin{figure}
\begin{center}
\includegraphics[width=0.45\textwidth,height=0.9\textheight]{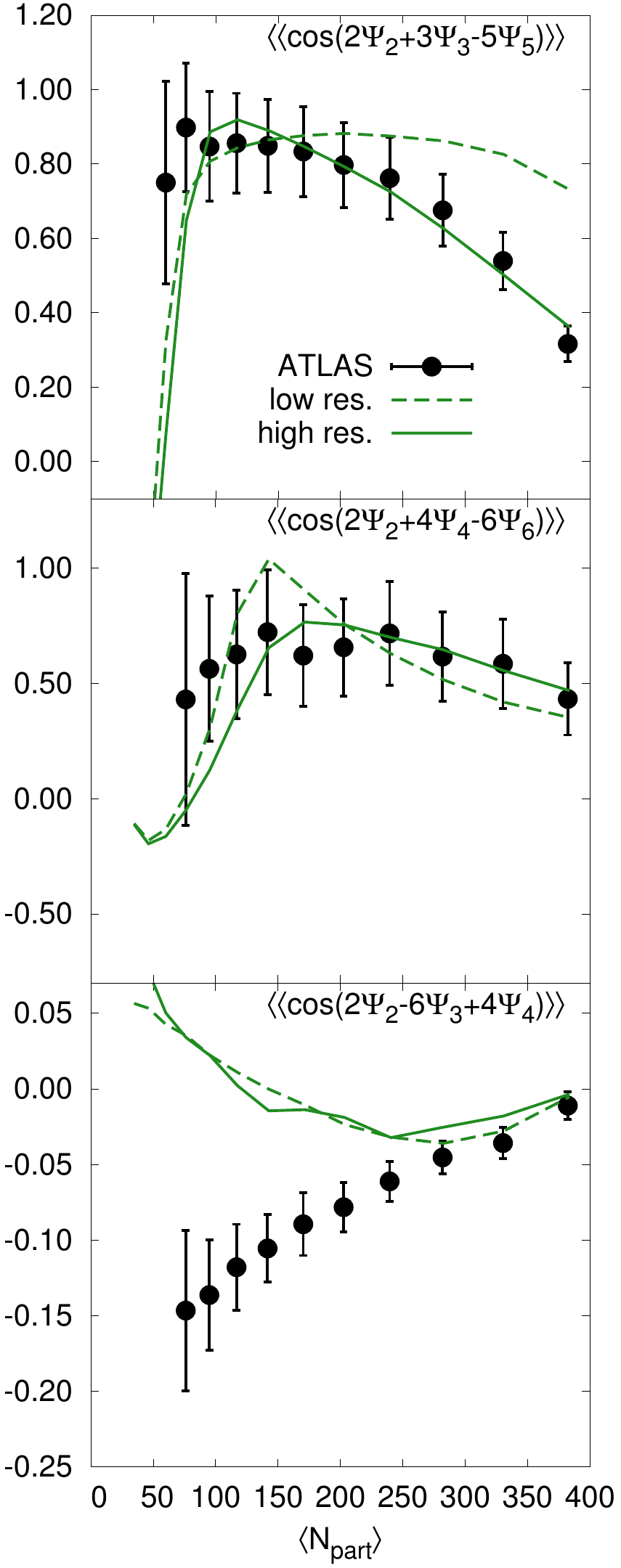} 
\includegraphics[width=0.45\textwidth,height=0.9\textheight]{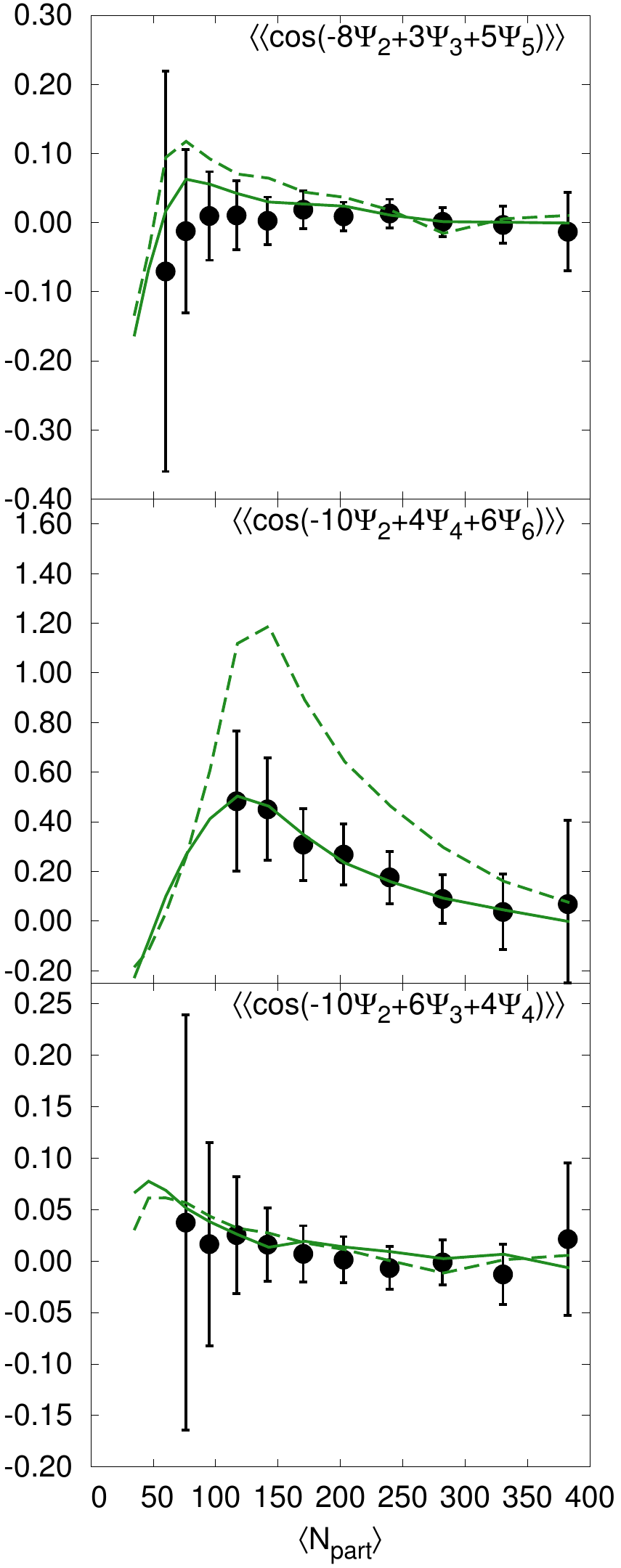} 
\caption{
Three plane correlations  using the non-linear response formalism. 
Here $\eta/s=1/4\pi$ for PHOBOS Monte-Carlo Glauber initial conditions.
The data are from the ATLAS collaboration \cite{ATLASCorrelations}.
The solid lines indicate the high resolution limit, \Eq{highres}, while the dashed
lines indicate the low resolution limit, \Eq{lowres}. 
\label{threeplanes} 
}
\end{center}
\end{figure}

\begin{figure}
\begin{center}
\includegraphics[width=0.45\textwidth,height=0.9\textheight]{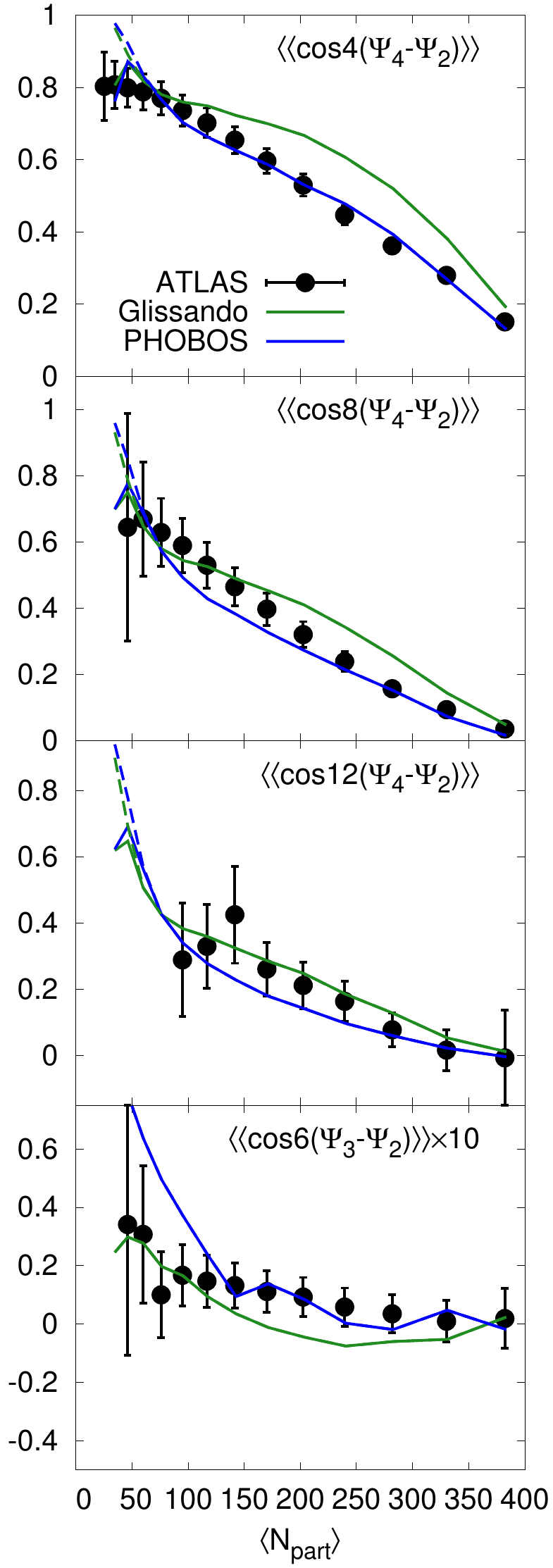} 
\includegraphics[width=0.45\textwidth,height=0.9\textheight]{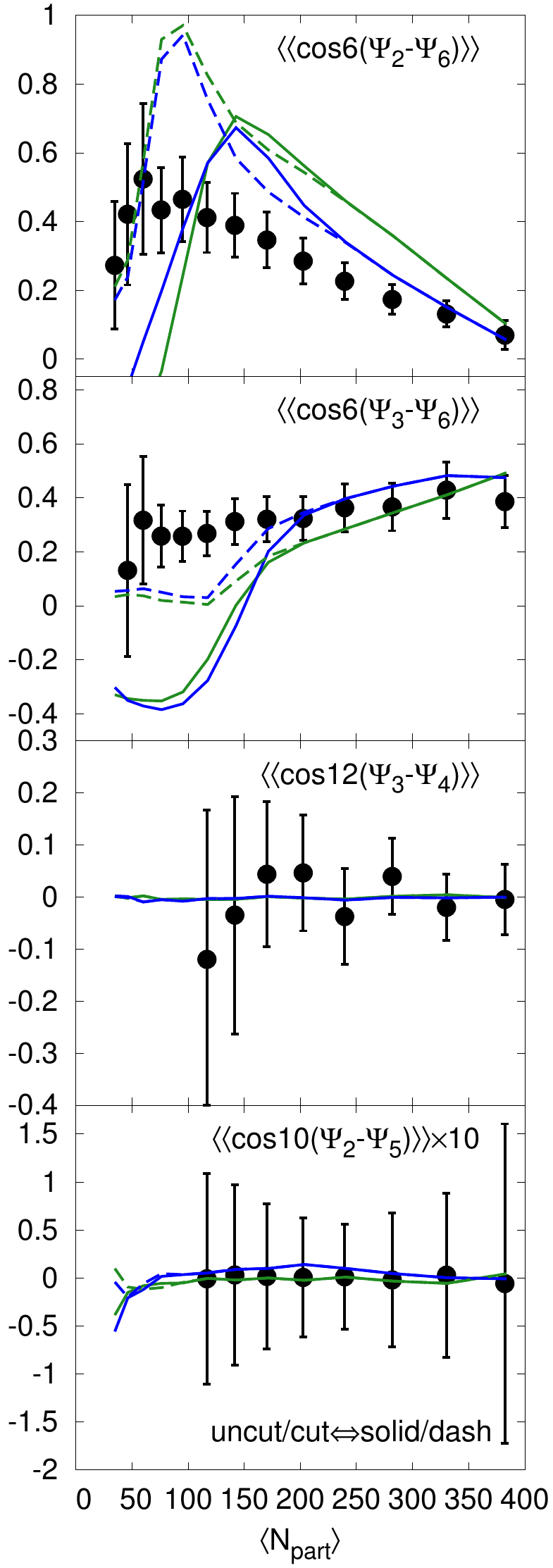} 
\caption{ 
   (Color online) A comparison of 
the two-plane correlations in the high resolution limit  for two different Glauber models, Glissando~\cite{Broniowski:2007nz} and the PHOBOS 
Glauber~\cite{Alver:2008aq}. 
The solid lines (un-cut) include the negative response in peripheral collisions due to a large $\delta f$, while
the dashed lines (cut) truncate the negative response -- see \Sect{diss}.
\label{2epc_compare}
}
\end{center}
\end{figure}

\begin{figure}
\begin{center}
\includegraphics[width=0.45\textwidth,height=0.9\textheight]{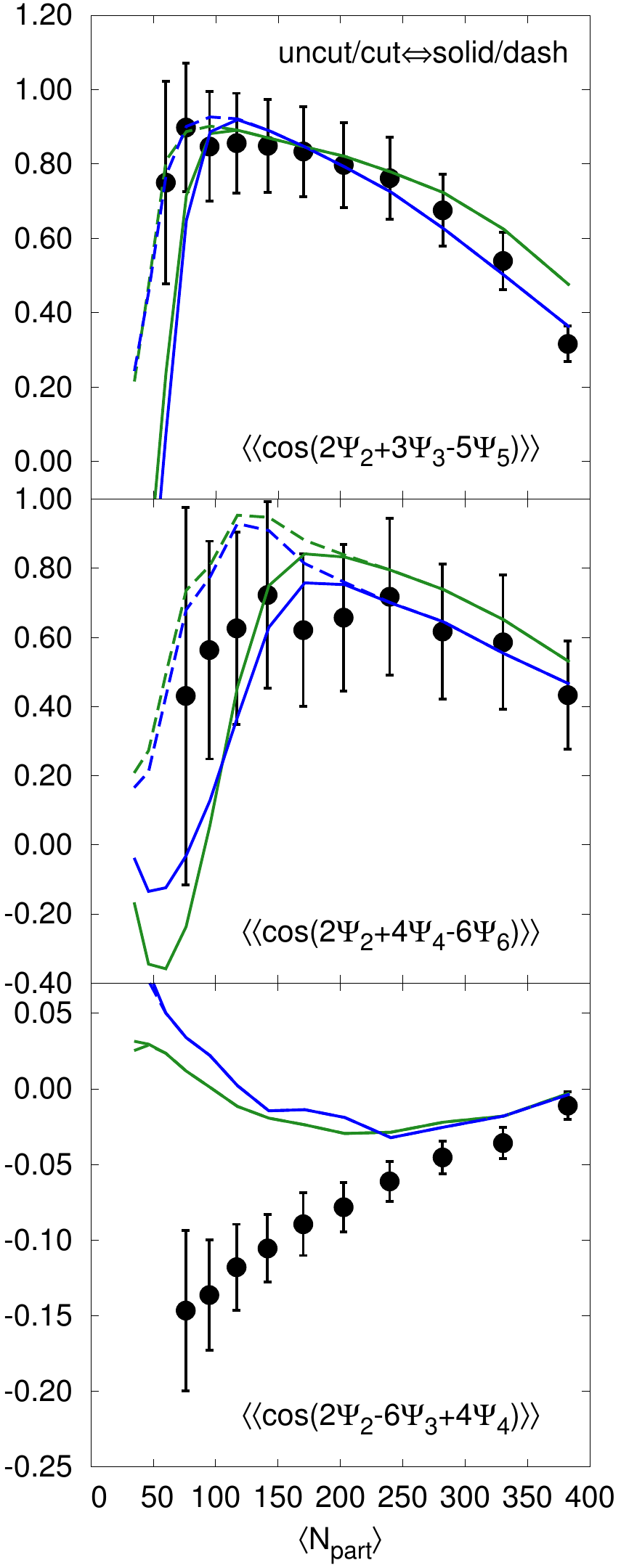} 
\includegraphics[width=0.45\textwidth,height=0.9\textheight]{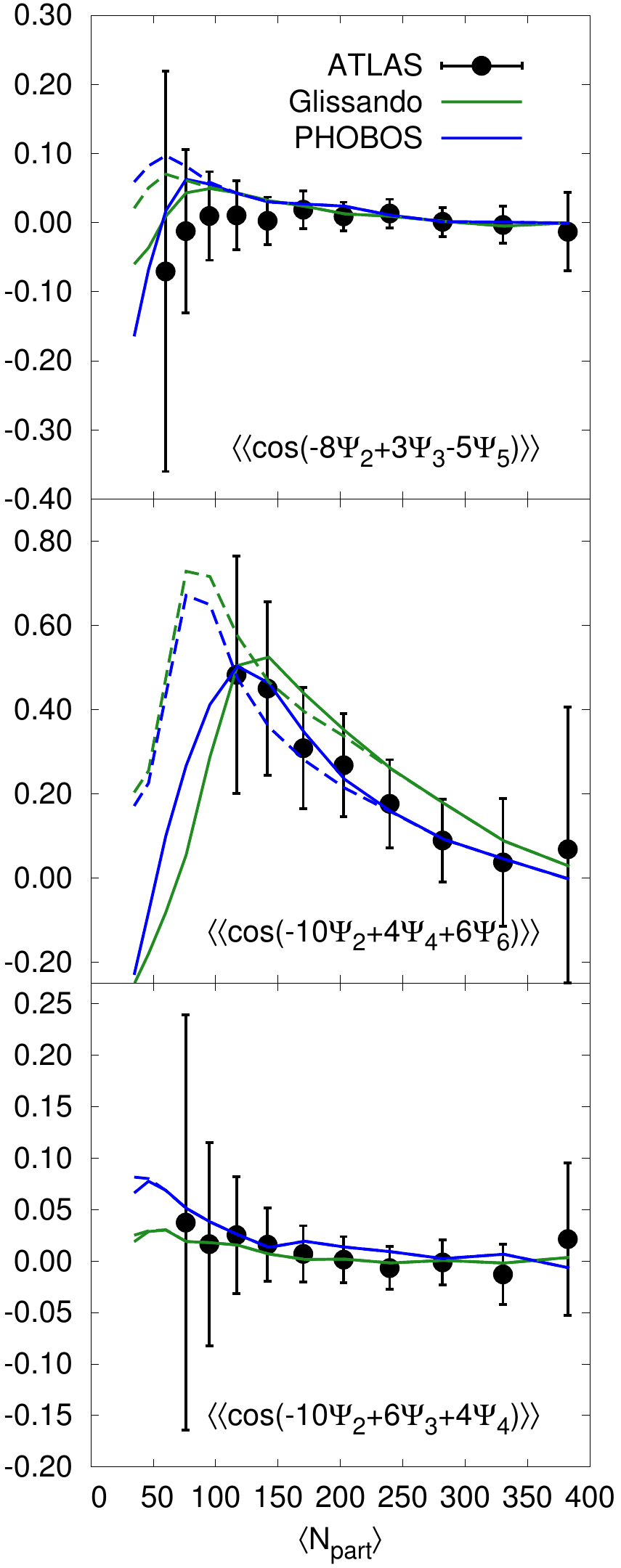} 
\caption{ 
   (Color online) A comparison of
the three-plane correlations in the high resolution limit for two different Glauber models, Glissando~\cite{Broniowski:2007nz} and the PHOBOS Glauber~\cite{Alver:2008aq}. 
The solid lines (un-cut) include the negative response in peripheral collisions due to a large $\delta f$, while
the dashed lines (cut) truncate the negative response -- see \Sect{diss}.
\label{3epc_compare}
}
\end{center}
\end{figure}

Figs.~\ref{twoplanes} and ~\ref{threeplanes} 
show a comparison
of the measured two and three plane correlation functions
with the response formalism in the  high and low resolution limits using 
the PHOBOS Glauber.
To test the sensitivity to the Glauber model in the high resolution limit
we compare two widely used monte-carlos -- the PHOBOS Monte Carlo Glauber \cite{Alver:2008aq} and
Glissando \cite{Broniowski:2007nz}. In Figs.~\ref{2epc_compare} and \ref{3epc_compare}, the predictions of viscous
hydrodynamics based on these two initial state models are shown by the blue and green lines, respectively.
The two Glauber models give similar results,
although the correlations from Glissando are somewhat stronger.

For the highest harmonics (such as $v_6$),  viscous corrections
in peripheral collisions can become too large to be trusted.
In this regime the linear and non-linear response coefficients can become
negative as a result of the first viscous correction to the distribution 
function~\cite{Teaney:2012ke}. 
Second order corrections to 
the viscous distribution are positive~\cite{Teaney:2013gca}, 
suggesting that such negative response coefficients are artificial.
Indeed, kinetic theory simulations have positive response coefficients
for all values of the Knudsen parameter~\cite{Alver:2010dn}.
To understand when viscous corrections  to the 
response coefficients are out of control, we have
performed two simulations. In the first case (un-cut), we blindly
allow the response coefficients to become negative. In
the second case (cut), we
set these coefficients to zero (as a function of centrality) when they turn negative.
In Figs.~\ref{2epc_compare} and \ref{3epc_compare} we show the correlation results
of the un-cut (solid) and cut (dashed) response  coefficients.
As seen in these figures, the ambiguity is noticeable
only for peripheral collisions, and for
correlations involving the highest harmonic, $\Psi_6$. 
Examining the $\dlangle\cos6(\Psi_6 - \Psi_3) \drangle$ correlation,
we see that the negative dive in peripheral collisions is 
an artifact of  out-of-control viscous corrections. A similar
negative dive is seen in event-by-event hydro simulations~\cite{Qiu:2012uy}.


Inspecting  these correlations, we make the following observations.
First, many of the most important correlation functions are reasonably
reproduced, at least if the high resolution limit is used.  
The agreement with the low resolution limit is not as good. The 
ambiguities in the measurement can be avoided by taking definite
moments as in \Eq{doneright}~\cite{Bhalerao:2011yg}.  Examining  the definitions of 
the high and low resolution limits (Eqs.~\ref{highres} and \ref{lowres}),  we see that the difference
between the two measurements can be best quantified by 
measuring the probability distribution $P(v_n)$~\cite{Aad:2013xma}, or the moments of this distribution~\cite{Alver:2007qw}, \emph{e.g.}  for $v_2$
\st
   (v_{2}\{2\})^2 \equiv \llangle v_2^2 \rrangle \quad \mbox{and} \quad  (v_2\{4\})^4 \equiv - \left[\llangle v_2^4  \rrangle - 2\llangle v_2^2 \rrangle \right] \, .
\stp
It is then  a separate and important question whether the 
response formalism outlined here can reproduce these probability distributions.
This will be addressed in future work.

There are a few correlations which are seemingly not well reproduced even 
in the high resolution limit. First, one could
hope for better agreement with the correlations involving $\Psi_6$ such
as $\cos(6\Psi_3 - 6\Psi_6)$ and $\cos(6\Psi_2 -6\Psi_6)$.  
$v_6$ is a relatively high harmonic, and viscous corrections
are not in perfect control in peripheral collisions~\cite{Teaney:2013gca}. 
This is clearly evident in 
\Fig{2epc_compare} which estimates the contributions of higher order 
viscous corrections to the distribution function (see above). For the $\Psi_6$-correlations (and no others),  these 
corrections are large in peripheral collisions. 

The most troubling  correlation function, 
which is not qualitatively reproduced by the response formulation,
is $\cos(2\Psi_2 - 6\Psi_3+ 4\Psi_4)$.  It is possible that
that this discrepancy stems from an underestimate of the mixing of 
$v_1$ with other modes, which naturally mixes $v_4$ with $v_3$.
Indeed, a preliminary analysis suggests that this correlation
is closely related to the transverse shift from the geometrical
center to the center of participants. The $2,3,4$ correlation is
qualitatively
reproduced by event-by-event hydrodynamics \cite{Qiu:2012uy}.

\begin{figure}
\includegraphics[width=0.49\textwidth]{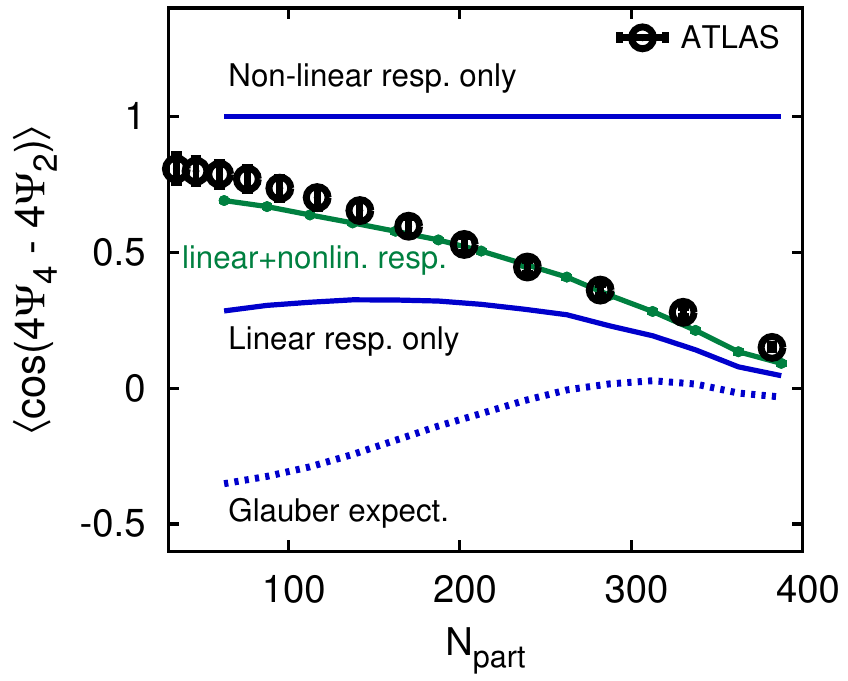}
\includegraphics[width=0.49\textwidth]{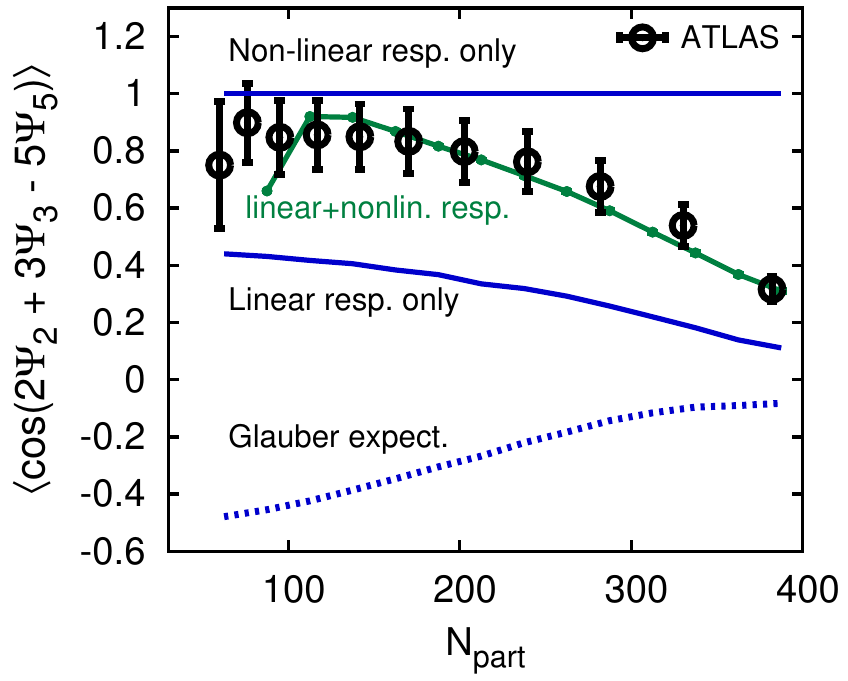}
\caption{ 
   The separate contributions of the linear and non-linear response to a two-plane correlation, $\llangle \cos(4\Psi_4 - 4\Psi_2) \rrangle$,  and 
   a three-plane correlation,  $\llangle \cos(2\Psi_2 + 3\Psi_3 - 5\Psi_5) \rrangle$. The dashed lines show the naive Glauber expectation (see text). The data
   is from \Ref{ATLASCorrelations}.
\label{linnonlin}
}
\end{figure}

It is important and instructive to understand the hydrodynamic origin of the correlations
presented in these figures. This is best understood by examining the 
linear and non-linear contributions separately. \Fig{linnonlin}(a) and (b) 
illustrate this decomposition with the correlations 
$\llangle \cos(4(\Psi_2 - \Psi_4) ) \rrangle$
and 
$\llangle \cos(2\Psi_2 + 3\Psi_3 - 5\Psi_5) \rrangle$  respectively.
For definiteness, we study the 2,3,5 combination shown in \Fig{linnonlin}(b). 
The naive 
expectation of the Glauber model 
(where $v_n$ is proportional to the $n-$th order
moment based eccentricity)
 is shown by the dotted line, and has 
the wrong sign. In the naive approach the observed correlation 
between the event plane angles $2,3,5$
arises from the correlations between the angles associated 
with the corresponding moment based eccentricities.

In the current work the $v_5$ is produced through a combination of 
the linear and non-linear response.  
\begin{itemize}
   \item In  linear response, $v_5$ is proportional 
to the $5$-th cumulant $\epsilon_5$, and the correlation between
the event plane angles  $\Psi_2,\Psi_3,\Psi_5$ reflects 
the initial state correlation between the associated cumulant angles, $\Phi_2,\Phi_3,\Phi_5$.   
The predictions of linear response are shown in \Fig{linnonlin}, 
and fail to reproduce the observed correlations
in non-central collisions.  
\item In non-linear response, $v_5$ is determined 
through the mode mixing of $v_2$ and $v_3$. If $v_5$ 
was determined entirely by this mechanism, the $\Psi_5$ 
 event plane would be entirely determined by $\Psi_2$ and $\Psi_3$,
 leading to a perfect $2,3,5$ correlation.
 This prediction of non-linear  response is also shown in \Fig{linnonlin}.
 \end{itemize}

In general, $v_5$ is determined by a weighted average of  the
linear and non-linear response curves. The relative size of
these two contributions is determined by viscous hydrodynamics
which predicts the magnitude of these response coefficients as a function 
of centrality.
Evidently, hydrodynamics and the response
formalism reproduces the centrality dependence of the observed correlation
functions.  It is  satisfying to see how the data transition
between the linear response curves in central collisions, and the
non-linear response curves in peripheral collisions.

Finally, we conclude by discussing the importance of higher 
order terms in the 
response formalism. First, we have neglected the third order mixing of 
harmonics. The most important third order term is proportional
to $\E_2^3$, and we have found that this term  is small
compared to the $\E_3^2$ and $\E_2 \E_4 $ terms.
Thus, the response formalism seems to converge,  and
including the mixing of higher harmonics will not change 
the results of this study  significantly.

In the future 
it will be important to characterize the fluctuations around
the response formalism. For any given initial state characterized
by a few macroscopic cumulants such as $\E_2,\E_3,\E_4 \ldots$, 
the observed $v_n$ will on \emph{average} be given by the response formalism.
However, additional fluctuations (which leave the macroscopic cumulants fixed) will reduce the perfect correlation between $v_2,v_3,v_4,\ldots$ and the
predictions of non-linear response.
Thus, in general, the response formalism will overestimate the strength
of the correlations that are observed. Ideally, the fluctuations
around the response formalism can be parametrized by universal Gaussian
noise, which will be independent of the microscopic details of the 
initial state. The study of fluctuations around the response 
formalism is left for future work.

\vspace{\baselineskip}
\noindent{ \bf Acknowledgments:} \\
{}\\
We thank J.~Y. Ollitrault, Z.~Qiu, U.~Heinz, J.~Jia, and S.~Mohapatra for many constructive and insightful comments.  D.~Teaney is a RIKEN-RBRC fellow.  This work is supported by the Department of Energy, DE-FG-02-08ER4154. 
Li Yan is also funded  by the European Research Council under the
Advanced Investigator Grant ERC-AD-267258.

\bibliography{pl_ref}

\end{document}